\documentclass[aip,jcp,reprint,onecolumn,frontmatterverbose,groupedaddress]{revtex4-1}
\usepackage{graphicx}
\usepackage{dcolumn}
\usepackage{bm}
\usepackage[mathlines]{lineno}
\usepackage{amssymb}

\newcommand{\sign}{\mathrm{sign}}
\newcommand{\M}{\mathrm{M}}

\begin{document}

\title[About linear superpositions of special exact solutions of Veselov-Novikov equation]{About linear superpositions of special exact solutions of Veselov-Novikov equation}

\author{V.G. Dubrovsky}
\email[E-mail: ]{dubrovsky@ngs.ru}
\author{A.V. Topovsky}
\email[E-mail: ]{topovsky.av@gmail.com}
\affiliation{Novosibirsk State Technical University, Karl Marx prosp. 20, Novosibirsk 630092, Russia.}

\date{\today}

\begin{abstract}
New exact solutions, nonstationary and stationary, of Veselov-Novikov (VN) equation in the forms of linear superpositions of arbitrary number of exact special solutions $ u^{(n)}$, $n=1,\ldots,N$  are constructed via $\overline \partial$-dressing method in such a way that the sums $u= u^{(k_1)}+\ldots+ u^{(k_m)}$, $1\leqslant k_1<k_2<\ldots<k_m\leqslant N$ of arbitrary subsets of these solutions are also exact solutions of VN equation. The presented linear superpositions include as superpositions of special line solitons with zero asymptotic values at infinity and also superpositions of special plane wave type singular periodic solutions. By construction these exact solutions represent also new exact transparent potentials of 2D stationary Schr\"{o}dinger equation and can serve as model potentials for electrons in  planar structures of modern electronics.
\end{abstract}

\pacs{02.30.Ik, 02.30.Jr, 02.30.Zz, 05.45.Yv}

\maketitle
\section{\label{Section_1}Introduction}
\label{Section_1}
\setcounter{equation}{0}
Among different (2+1)-dimensional integrable nonlinear equations \cite{NovZakh_BookSoliton, AblClark_BookSoliton, KonopBook_92, KonopBook_93, Manakov_81, ZakhManak_85, ZakharovProc_88, BogdManak_88, FokasZakh_92} prominent place takes the famous Veselov-Novikov (VN) equation \cite{Nizhnik_80,VeselovNovikov_84}:
\begin{equation}\label{VN}
u_{t} + \kappa u_{zzz} + \overline \kappa u_{\bar z\bar z\bar z} +
  3\kappa(u\partial^{-1}_{\bar z}u_{z})_{z} +
  3\bar\kappa(u\partial^{-1}_{z}u_{\bar z})_{\bar z}=0
\end{equation}
where $u(z,\bar z,t)$ is scalar function, $\kappa$ is some complex  constant;
$z=x+iy$, $\bar z=x-iy$;
$\partial^{-1}_{z}$ and $\partial^{-1}_{\bar z}$ are operators inverse to
$\partial_{z}$ and $\partial_{\bar z}$,
$\partial^{-1}_{\bar z}\partial_{\bar z}=\partial^{-1}_{z}\partial_{z}=1$.

VN equation can be represented as compatibility condition in the form of Manakov's triad \cite{ManakovTriads}:
\begin{equation}\label{VN dressing triada Manakova}
[L_{1}, L_{2}] = BL_{1},\quad
  B =3\big(\kappa\partial_{\bar z}^{-1}u_{zz}+
  \overline\kappa\partial_{z}^{-1}u_{\bar z\bar z}\big)
\end{equation}
of two linear auxiliary problems
\begin{equation}\label{VN dressing auxiliary problems}
L_{1}\psi = \big(\partial_{z\bar z}^{2}+u\big)\psi=0,
\end{equation}
\begin{equation}\label{VN dressing auxiliary problems2}
L_{2}\psi = \big(\partial_{t}+\kappa\partial_{z}^{3}+\overline\kappa\partial_{\bar z}^{3}+
      3\kappa\big(\partial_{\bar z}^{-1}u_{z}\big)\partial_{z}+
      3\overline\kappa\big(\partial_{z}^{-1}u_{\bar z}\big)\partial_{\bar z}\big)\psi=0.
   \end{equation}

Several classes of exact solutions of VN equation (\ref{VN}) have been constructed in last three decades (1980 -- 2010) via different methods
~\cite{Nizhnik_80,VeselovNovikov_84,GrinevManak_86,GrinevManak_88,GrinevNovikov_88,MatveevSalle_Book,Grinevich_Thesis,
Grinevich_00, AthorneNimmo_91,DubrForm_01,DubrForm_03, DubrTopBasTMP,DubrTopBas_arxiv,DubrTopBasTMP1}, see also the books~\cite{KonopBook_92,KonopBook_93}.
These solutions include finite-zone type solutions ~\cite{VeselovNovikov_84}, rationally localized solutions~\cite{GrinevManak_86,
GrinevManak_88,GrinevNovikov_88,Grinevich_Thesis,
Grinevich_00, DubrForm_01,DubrForm_03} or lumps,  solutions with functional parameters \cite{Nizhnik_80,MatveevSalle_Book,DubrTopBasTMP,DubrTopBasTMP1}, multi-line soliton
solutions~\cite{DubrTopBas_arxiv, DubrTopBasTMP1}  and so on. Underline that the first auxiliary linear problem (\ref{VN dressing auxiliary problems}) is nothing but the 2D stationary Schr\"{o}dinger equation so exact solutions of VN equation constructed via all IST approaches are also transparent potentials of this Schr\"{o}dinger equation.

Recently in the paper~\cite{DubrTopBasTMP} the class of exact solutions with functional parameters and constant asymptotic values $-\epsilon$ at infinity
\begin{equation}\label{VN dressing asymptotic value of u}
  u(z,\bar z,t)=\tilde{u}(z,\bar z,t)-\epsilon,\quad \tilde{u}(z,\bar z,t)|_
  {|z|\rightarrow\infty}\rightarrow 0
\end{equation}
of VN equation (\ref{VN}) via $\overline \partial$-dressing method of Zakharov and Manakov \cite{ZakhManak_85, ZakharovProc_88, BogdManak_88, FokasZakh_92}  has been calculated and subclass of multi-line soliton solutions has been presented~\cite{DubrTopBasTMP}.

In another paper~\cite{DubrTopBasTMP1} (see
also~\cite{DubrTopBas_arxiv}) it was established that for some special solutions $ u^{(1)}$ and $ u^{(2)}$, i. e. for special linear (plane) solitons or for special plane wave type singular periodic solutions, with zero  value $E=-2\epsilon=0$ of energy level of corresponding 2D stationary Schr\"{o}dinger equation, their sum $ u^{(1)}+ u^{(2)}$ is also exact solution of VN equation.

In present paper this result~\cite{DubrTopBasTMP1,DubrTopBas_arxiv}  is generalized to the case of linear superpositions of arbitrary number of special line solitons (or special plane wave type periodic solutions) $u^{(n)}$, $n=1,\ldots,N$ in such a way, that the sums of arbitrary subsets of these solutions
\begin{equation}\label{arbitrary subsets}
u=u^{(k_1)}+\ldots+u^{(k_m)}, \quad 1\leqslant k_1<k_2<\ldots<k_m\leqslant N
\end{equation}
are also exact solutions of VN equation (\ref{VN}).

For convenience here some useful formulas of  $\overline \partial$-dressing method for VN equation (\ref{VN})
~\cite{KonopBook_92,KonopBook_93,DubrForm_01,DubrForm_03,DubrTopBasTMP,DubrTopBasTMP1,DubrTopBas_arxiv} are presented. Central object of this method is the scalar wave function $\chi(\lambda;z,\bar z,t)$
\begin{equation}\label{WaveFunction}
\chi(\lambda;z,\bar z,t)=e^{-F(\lambda;z,\bar z,t)}\psi(z,\bar z,t),\quad F(\lambda;z,\bar z,t)=i\big[\lambda z-\frac{\epsilon}{\lambda}\bar z+\big(\kappa\lambda^{3}-\overline\kappa
\frac{\epsilon^{3}}{\lambda^{3}}\big)t\big]
\end{equation}
 which satisfies to corresponding $\overline \partial$-problem or equivalently to following singular integral equation:
\begin{equation}\label{di_problem1}
\chi (\lambda) = 1 + \int\int\limits_C {\frac{d{\lambda }'\wedge
d{\overline {\lambda'}}}{2\pi i(\lambda'-\lambda)}}
\int\int\limits_C  \chi(\mu,\overline{\mu})
R(\mu ,\overline \mu ;\lambda'
,\overline {\lambda' }){d\mu \wedge d\overline{\mu }}.
\end{equation}
Here canonical normalization $\chi\rightarrow \chi_\infty=1$ as $\lambda\rightarrow\infty$ of wave function is assumed and the kernel $R$ is given by the formula~\cite{KonopBook_93,DubrTopBasTMP,DubrTopBasTMP1,DubrTopBas_arxiv}
\begin{equation}
R(\mu ,\overline \mu ;\lambda,\bar {\lambda };z,\bar z,t)
  =R_0 (\mu ,\overline \mu ;\lambda,\bar {\lambda })
  e^{F(\mu; z,\bar z,t)-F(\lambda; z,\bar z,t)}.
\end{equation}

Solutions $u(z,\bar z,t)$ of VN equation  are expressed via reconstruction formulas \cite{KonopBook_93,DubrForm_01,DubrForm_03,DubrTopBasTMP,DubrTopBasTMP1,DubrTopBas_arxiv}:
\begin{equation}\label{u_reconstructFormulae}
  u= -\epsilon+i\epsilon\chi_{1z}=-\epsilon-i\chi_{-1\bar z}
\end{equation}
through the coefficients $\chi_{1}$ and $\chi_{-1}$ of Taylor's series
\begin{equation}\label{series of chi}
\chi =\chi_{0}+\chi_{1}\lambda+\chi_{2}\lambda^{2}+\ldots, \quad
  \chi =\chi_{\infty}+\frac{\chi_{-1}}{\lambda}+\frac{\chi_{-2}}{\lambda^{2}}+\ldots
\end{equation}
expansions in the neighborhoods of points $\lambda=0$ and $\lambda=\infty$ of complex plane $\mathbb{C}$.
In constructing of exact solutions $u$ of VN equation  (\ref{VN}) two conditions must be satisfied ~\cite{KonopBook_93,DubrForm_01,DubrForm_03,DubrTopBasTMP,DubrTopBasTMP1,DubrTopBas_arxiv}: the condition of potentiality  of operator $L_1$, or the absence in  the first auxiliary linear problem
(\ref{VN dressing auxiliary problems}) of the terms with first derivatives $u_1\partial_z$ and $u_2\partial_{\bar z}$, and the condition of reality $u=\bar u$ of solutions.

The potentiality condition on operator $L_1$ in (\ref{VN dressing auxiliary problems}), or equivalently in terms of wave function $\chi$ the condition \linebreak $\chi_0=1$~\cite{KonopBook_93,DubrForm_01,DubrForm_03,DubrTopBasTMP,DubrTopBasTMP1,DubrTopBas_arxiv},  imposes severe restrictions on the kernel $R_0$ of $\overline\partial$-problem. The condition of reality of solutions $u=\bar u$ leads to another following restriction on the kernel
$R_0$~\cite{DubrForm_01,DubrForm_03,DubrTopBasTMP,DubrTopBasTMP1,DubrTopBas_arxiv}:
\begin{equation}\label{real_condition_VN}
  R_0(\mu,\overline{\mu};\lambda,\overline{\lambda})
  =\frac{\epsilon^3}{{|\mu|}^2{|\lambda|}^2
  \overline{\mu}\overline{\lambda}}\;
  \overline{R_0\left(-\frac{\epsilon}{\overline{\lambda}},
  -\frac{\epsilon}{\lambda}, -\frac{\epsilon}{\overline{\mu}}
  -\frac{\epsilon}{\mu}\right)},
\end{equation}
this restriction was obtained in the limit of \,"weak"\, fields.
Both conditions were successfully applied in calculations of broad classes of exact solutions of VN equation (\ref{VN}) such as lumps~\cite{DubrForm_01,DubrForm_03}, solutions with functional parameters, multi-line solitons and plane wave type singular periodic solutions~\cite{DubrTopBasTMP,DubrTopBasTMP1,DubrTopBas_arxiv}.

In the present note we do not use the limit of weak fields and impose the reality condition $u=\overline u$ directly to calculated complex solutions satisfying only to potentiality condition. This approach makes it possible to receive besides multi-line soliton solutions also plane wave type singular periodic solutions (this was shown at first in ~\cite{DubrTopBas_arxiv,DubrTopBasTMP1})  and their's superpositions.

By the application of $\overline\partial$-dressing in the special limit of  zero energy level we obtain in present paper new exact solutions, nonstationary and stationary, of VN equation in the forms of linear superpositions of special line solitons  and also linear superpositions of special plane wave type singular periodic solutions. By construction these exact solutions represent also new exact transparent potentials of 2D stationary Schr\"{o}dinger equation and can find an applications as model potentials for electrons in  planar structures of modern electronics.

\section{Nonlinear superpositions of complex solutions of VN equation}
\label{Section_2}
The choice of delta-functional kernel
\begin{equation}\label{delta_kernel_sec2}
  R_0(\mu,\bar{\mu};\lambda,\bar{\lambda})=\pi\sum\limits_n A_n
  \delta(\mu-\M_n)\delta(\lambda-\Lambda_n)
\end{equation}
with complex constant  coefficients $A_n$ and complex discrete spectral parameters $\M_n\neq\Lambda_n$ leads to simple determinant formula~\cite{DubrTopBasTMP1,DubrTopBas_arxiv}
\begin{equation}\label{Solution_general_formula_sec2}
u=-\epsilon+\frac{\partial^2}{\partial z\partial\bar z}\ln\det A,\quad
A_{lk}=\delta_{lk}+ \frac{2iA_{k}}{\M_{l}-\Lambda_{k}}e^{F(\M_{l})-F(\Lambda_{k})}
\end{equation}
for exact multi-line soliton and plane wave type singular periodic solutions of VN equation. The main problem in using this formula is satisfaction to reality and potentiality conditions.

It was shown in~\cite{DubrTopBasTMP1,DubrTopBas_arxiv} that the choice of kernel $R_0$ (\ref{delta_kernel_sec2}) in the form
\begin{eqnarray}\label{sum delta_kernel_satisfied_potent}
  R_0(\mu,\bar{\mu};\lambda,\bar{\lambda})=\pi\sum\limits_{n=1}^{N}\Big[a_n\lambda_n\delta(\mu-\mu_n)
  \delta(\lambda-\lambda_n) + a_n\mu_n\delta(\mu+\lambda_n)\delta(\lambda+\mu_n)\Big]
\end{eqnarray}
of $N$ paired terms with discrete spectral parameters $(\mu_n,\lambda_n)$ allows to satisfy the potentiality condition $\chi_0=1$.
In the simplest cases $N=1,2$ one obtains from  (\ref{delta_kernel_sec2}) -- (\ref{sum delta_kernel_satisfied_potent}) the following expressions for $\det A$~\cite{DubrTopBasTMP1,DubrTopBas_arxiv}:
\begin{equation}\label{detA N=1}
N=1: \det A =
\Big(1+s_1e^{\Delta F(\mu_1,\lambda_1)}\Big)^2,
\end{equation}
\begin{equation}\label{detA N=2}
 N=2:\det A = \Big(1+s_1e^{\Delta F(\mu_1,\lambda_1)}+
 s_ne^{\Delta F(\mu_n,\lambda_n)}+we^{\Delta F(\mu_1,\lambda_1)+\Delta F(\mu_n,\lambda_n)}\Big)^2.
\end{equation}
Here, in (\ref{sum delta_kernel_satisfied_potent}) -- (\ref{detA N=2}) $a_n$, $\mu_n$, $\lambda_n$ $(n=1,\ldots,N)$ are some complex constants; $\mu_n$ and $\lambda_n$ also known as discrete spectral parameters which give spectral characterization for corresponding exact solutions.
The quantities $s_n$, $w$ and $\Delta F(\mu_n,\lambda_n)$  are given by the formulas:
\begin{equation}\label{s_k,Delta F}
 s_n :=ia_n\frac{\mu_n + \lambda_n}{\mu_n - \lambda_n}; \quad
 \Delta F(\mu_n,\lambda_n):=F(\mu_n) -  F(\lambda_n),
\end{equation}
\begin{equation}\label{w}
w:=-s_1 s_n \cdot \frac{(\lambda_{1}-\lambda_{n})(\lambda_{n}+\mu_{1})(\mu_{1}-\mu_{n})(\lambda_{1}+\mu_{n})}
{(\lambda_{1}+\lambda_{n})(\lambda_{n}-\mu_{1})(\mu_{1}+\mu_{n})(\lambda_{1}-\mu_{n})}.
\end{equation}
The expression for $\det A$ in the case $N=2$ (\ref{detA N=2}) is generated by two pairs of terms in (\ref{sum delta_kernel_satisfied_potent}) with discrete spectral variables $(\mu_1,\lambda_1)$ and $(\mu_n,\lambda_n)$.

The formula for generally complex solution corresponding to one arbitrary pair $(\mu_m,\lambda_m),  (m=1,\ldots,N)$ of discrete spectral variables due to (\ref{delta_kernel_sec2}) -- (\ref{detA N=1}) and (\ref{s_k,Delta F}) has the form:
\begin{equation}\label{uN=1Gen}
 u^{(m)}(z,\bar z,t)=-\epsilon+\tilde{u}^{(m)}(z,\bar z,t)=
-\epsilon-\epsilon{\frac{2s_m(\mu_m-\lambda_m)^2}{\mu_m\lambda_m}}
{\frac{e^{\Delta F(\mu_m,\lambda_m)}}
{(1+s_m e^{\Delta F(\mu_m,\lambda_m)})^2}}.
\end{equation}

It is remarkable that for  $w=s_1s_n$ in (\ref{detA N=2}) for case $N=2$ of two pairs of terms in (\ref{sum delta_kernel_satisfied_potent}) with spectral variables $(\mu_1,\lambda_1)$ and $(\mu_n,\lambda_n), (n=2,\ldots,N)$, i. e. for the choice
\begin{equation}\label{SeparatCondition0}
\frac{(\lambda_{1}-\lambda_{n})(\lambda_{n}+\mu_{1})(\mu_{1}-\mu_{n})(\lambda_{1}+\mu_{n})}
{(\lambda_{1}+\lambda_{n})(\lambda_{n}-\mu_{1})(\mu_{1}+\mu_{n})(\lambda_{1}-\mu_{n})}=-1,
\end{equation}
which is equivalent to relation
\begin{equation}\label{SeparatCondition}
(\lambda_1-\mu_1)(\lambda_n-\mu_n)(\lambda_1\mu_1+\lambda_n\mu_n)=0, \quad n\neq1,
\end{equation}
the expression for $\det A$ (\ref{detA N=2}) greatly  simplifies
\begin{equation}\label{detA N=2SimplForm}
\det A = \left(1+s_1e^{\Delta F(\mu_1,\lambda_1)}\right)^{2}
\left(1+s_n e^{\Delta F(\mu_n,\lambda_n)}\right)^{2}.
\end{equation}
The solutions $\mu_1=\lambda_1$ and $\mu_n=\lambda_n$ of (\ref{SeparatCondition}) correspond to lumps (rationally decreasing at infinity exact solutions $u$~\cite{DubrForm_01,DubrForm_03} of VN equation) and in accordance with $M_n\neq\Lambda_n$ in (\ref{delta_kernel_sec2}) will not be considered here, so  it is assumed   below that $\mu_m\neq\lambda_m$ in (\ref{sum delta_kernel_satisfied_potent}) for all terms $m=1,\ldots,N$ with discrete spectral variables $\mu_m$, $\lambda_m$. Under this requirement the relations  (\ref{SeparatCondition0}), (\ref{SeparatCondition}) reduce to more simple ones:
\begin{equation}\label{SplittingCondition1}
\lambda_n\mu_n+\lambda_1\mu_1=0, \quad n=2,\ldots,N.
\end{equation}

An application of general formulas (\ref{delta_kernel_sec2}) -- (\ref{detA N=2}) in the case $N=2$ due to (\ref{SeparatCondition0}) or (\ref{SplittingCondition1}) leads to very simple expression for complex solution of VN equation
\begin{equation}\label{uTwoSolGen}
u(z,\bar z,t) = -\epsilon-2\epsilon\sum\limits_{m=1,n}
\frac{s_m(\mu_m-\lambda_m)^2}{\mu_m\lambda_m}
\frac{e^{\Delta F(\mu_m,\lambda_m)}}
{(1+s_m e^{\Delta F(\mu_m,\lambda_m)})^2}
\end{equation}
which is nonlinear superposition  $u=\epsilon+u^{(1)}+u^{(n)}$  of two solutions $u^{(1)}$ and $u^{(n)}$ of the type (\ref{uN=1Gen}) with corresponding   pairs of spectral variables $(\mu_1,\lambda_1)$ and $(\mu_n,\lambda_n)$.
Up to the constant $\epsilon$  the solution (\ref{uTwoSolGen}) is   the sum of complex solutions $u^{(1)}$ and $u^{(n)}$.

It is easy to prove also  that nonlinear superposition
\begin{eqnarray}\label{MultiSolitonGF}
 u(z,\bar z,t)=-\epsilon+\tilde u^{(1)}(z,\bar z,t)+
\sum\limits_{m=2}^{N}\tilde u^{(m)}(z,\bar z)
=-\epsilon-2\epsilon{\frac{s_1(\mu_1-\lambda_1)^2}{\mu_1\lambda_1}}
{\frac{e^{\Delta F(\mu_1,\lambda_1)}}
{(1+s_1 e^{\Delta F(\mu_1,\lambda_1)})^2}}-\nonumber \\
-2\epsilon\sum\limits_{m=2}^{N}
{\frac{s_m(\mu_m-\lambda_m)^2}{\mu_m\lambda_m}}
{\frac{e^{\Delta F(\mu_m,\lambda_m)}}
{(1+s_m e^{\Delta F(\mu_m,\lambda_m)})^2}}
\end{eqnarray}
of arbitrary number $N\geqslant2$ of complex solutions, solution $u^{(1)}(z,\bar z,t)=-\epsilon+\tilde u^{(1)}(z,\bar z,t)$ and $N-1\geqslant1$  solutions
$u^{(m)}(z,\bar z)=-\epsilon+\tilde u^{(m)}(z,\bar z)$, $m=2,\ldots,N$ of the type (\ref{uN=1Gen}) is also exact solution of VN equation, when conditions (\ref{SplittingCondition1}) are fulfilled and parameters $\mu_1$ and $\lambda_1$ are satisfied to additional restriction
\begin{equation}\label{SplittingCondition2}
\kappa\lambda_1^3-\overline\kappa\frac{\epsilon^3}{\mu_1^3}=0.
\end{equation}

Due to conditions (\ref{SplittingCondition1}), (\ref{SplittingCondition2}) the phases $\Delta F(\mu_m,\lambda_m)$ (\ref{s_k,Delta F}) in (\ref{MultiSolitonGF}) take the forms:
\begin{equation}\label{Phazes1NsolitonGF}
 \varphi_1(z,\bar z,t):=\Delta F(\mu_1,\lambda_1)=
i\left[(\mu_1 - \lambda_1)z -\left(\frac{\epsilon}{\mu_1}-\frac{\epsilon}{\lambda_1}\right)\bar z
-2\left(\kappa\lambda_1^3-\overline\kappa\frac{\epsilon^3}{\lambda_1^3}\right)t\right],
\end{equation}
\begin{equation}\label{PhazesnNsolitonGF}
 \varphi_{m}(z,\bar z):=\Delta F(\mu_m,\lambda_m)=
i\left[(\mu_m - \lambda_m)z -\left(\frac{\epsilon}{\mu_m}-\frac{\epsilon}{\lambda_m}\right)\bar z\right], \quad m=2,\ldots,N.
\end{equation}
These expressions (\ref{Phazes1NsolitonGF}) and (\ref{PhazesnNsolitonGF}) mean that the first complex solution $u^{(1)}(z,\bar z,t)=-\epsilon+\tilde u^{(1)}(z,\bar z,t)$ of superposition (\ref{MultiSolitonGF})
propagates with nonzero velocity in the plane $(x,y)$
but all other complex solutions
$u^{(m)}(z,\bar z)=-\epsilon+\tilde u^{(m)}(z,\bar z)$, $(m=2,\ldots,N)$
of superposition (\ref{MultiSolitonGF}) with $N\geqslant2$ are fixed
in the plane $(x,y)$ stationary solutions of VN equation (\ref{VN}).

One can prove also that every subsum of arbitrary terms $1\leqslant i<i+1<\ldots<j-1<j\leqslant N$ of sum (\ref{MultiSolitonGF})
\begin{equation}\label{SubSystSuperp}
u=-\epsilon+
\sum\limits_{n=i}^{j}\tilde{u}^{(n)}
\end{equation}
under conditions (\ref{SplittingCondition1}) and (\ref{SplittingCondition2}) is exact solution of VN equation.
Complex solutions of VN equation given by (\ref{SubSystSuperp})
due to (\ref{MultiSolitonGF}) and (\ref{Phazes1NsolitonGF}),
(\ref{PhazesnNsolitonGF})
can be divided on two classes: the class of nonstationary solutions with $i\geqslant 1$ and class of stationary solutions with $i\geqslant 2$.

\section{Linear superpositions of line soliton solutions  for Veselov-Novikov equation}

For construction of real multi-line solitons via (\ref{Solution_general_formula_sec2}) besides potentiality condition satisfied by the kernel $R_0$ of the type (\ref{sum delta_kernel_satisfied_potent}) the reality condition $u=\overline u$ for solutions $u$ must be fulfilled. This can be done choosing appropriately complex constants $a_n$ and complex discrete spectral parameters $(\mu_n,\lambda_n)$ in
(\ref{sum delta_kernel_satisfied_potent}) -- (\ref{SubSystSuperp}) by several ways~\cite{DubrTopBasTMP1,DubrTopBas_arxiv}. For example,
by imposing reality condition $u=\bar{u}$ on complex solutions (\ref{uN=1Gen}), (\ref{uTwoSolGen}), (\ref{MultiSolitonGF}) and (\ref{SubSystSuperp}) with additional assumption of real phases $\Delta F(\mu_n,\lambda_n)= \overline{\Delta F(\mu_n,\lambda_n)}$   (\ref{s_k,Delta F}) we have calculated real multi-line soliton solutions.

It was shown in the papers~\cite{DubrTopBasTMP1,DubrTopBas_arxiv} that to such real solutions $u$ leads the following choice of parameters
\begin{equation}\label{[1,2,0]VNSolitons}
a_n = -\overline{a}_n := ia_{n0},\quad  \mu_n = -{\frac{\epsilon}{\overline{\lambda}_n}}, \quad n=1,\ldots,N
\end{equation}
with real constants $a_{n0}$.
Due to (\ref{[1,2,0]VNSolitons}) and under additional assumption of positive values of real constants $s_n$ given by (\ref{s_k,Delta F})
\begin{equation}\label{p_kVNElliptic1}
s_n= a_{n0}\frac{\lambda_{n} +\mu_n}{\lambda_{n}-\mu_n}\stackrel{\mathrm{def}}{=}e^{\phi_{0n}} > 0, \quad n=1,\ldots,N
\end{equation}
the solution (\ref{uN=1Gen}) corresponding to one arbitrary pair $(\mu_n,\lambda_n)$,  $(n=1,\ldots,N)$ of discrete spectral variables takes the form of real nonsingular one-line soliton solution:
\begin{equation}\label{VN solution 1 u}
u^{(n)}(x,y,t)= -\epsilon +\tilde u^{(n)}(x,y,t)= -\epsilon +
\frac{|\lambda_n - \mu_n|^2}{2\cosh^2\frac{\varphi_n(x,y,t) + \phi_{0n}}{2}}
\end{equation}
where real phases $\varphi_n(x,y,t) :=\Delta F(\mu_n,\lambda_n)$ (\ref{s_k,Delta F}) due to (\ref{[1,2,0]VNSolitons}) have the form
\begin{equation}\label{VN two-solution 1 varphi1}
 \varphi_n(x,y,t) = 2|\lambda_n|\left(1+\frac{\epsilon}{|\lambda_n|^2}\right)(\vec{N}_n\vec{r}-V_n t), \quad n=1,\ldots,N,
\end{equation}
here $\vec{r}=(x,y)$,  $\vec{N}_n$  are unit vectors of normals to lines of constant values of phases $\varphi_{n}(x,y,t)$ and $V_n$ are corresponding velocities of one-line solitons
\begin{equation}\label{VN two-solution 1 varphi1Normals}
\vec{N}_n=\left(\frac{\lambda_{nI}}{|\lambda_{n}|},\frac{\lambda_{nR}}{|\lambda_{n}|}\right),\quad
V_{n}=-\frac{1}{|\lambda_n|}
\left(1+\frac{\epsilon\left(\epsilon-|\lambda_n|^2\right)}{|\lambda_n|^4}\right)\mathrm{Im}(\kappa\lambda^3_n),
\quad n=1,\ldots,N.
\end{equation}

For the cases of  nonlinear superpositions (\ref{uTwoSolGen}) $(N=2)$, (\ref{MultiSolitonGF}) and (\ref{SubSystSuperp}) $(N\geqslant2)$ of exact solutions of the type  (\ref{VN solution 1 u}) the conditions (\ref{SplittingCondition1}) for discrete spectral parameters $(\mu_n,\lambda_n)$, $(n>1)$
due to (\ref{[1,2,0]VNSolitons}) lead to following parametrization of $(\mu_n,\lambda_n)$
\begin{equation}\label{Parametrization}
\mu_n=i\tau_{n}^{-1}\mu_1, \quad \lambda_n=i\tau_{n}\lambda_1,\quad n=2,\ldots,N
\end{equation}
with arbitrary real constants $\tau_n$.
Nonsingular two-line soliton solution characterized by two pairs of discrete spectral variables $(\mu_1,\lambda_1)$ and $(\mu_2,\lambda_2)$ due to (\ref{uTwoSolGen}), (\ref{[1,2,0]VNSolitons}), (\ref{p_kVNElliptic1}) and (\ref{Parametrization}) takes the form
\begin{equation}\label{VN solution 1 u simple}
 u(x,y,t)= -\epsilon+\sum\limits_{n=1}^{2}\tilde u^{(n)}(x,y,t)=-\epsilon+\sum\limits_{n=1}^{2}
\frac{|\lambda_n - \mu_n|^2}{2\cosh^2{\frac{\varphi_{n}(x,y,t) + \phi_{0n}}{2}}},
\end{equation}
where $u^{(n)}(x,y,t)= -\epsilon+\tilde u^{(n)}(x,y,t)$, $(n=1,2)$ are one-line soliton solutions of the type
(\ref{VN solution 1 u}), the phases $\varphi_{n}(x,y,t),(n=1,2)$ under conditions (\ref{Parametrization}) are given by (\ref{VN two-solution 1 varphi1}).
Due to expressions for vectors of normals (\ref{VN two-solution 1 varphi1Normals}) and parametrization (\ref{Parametrization}) it is evident that solitons $u^{(1)}$ and $u^{(2)}$ of superposition
(\ref{VN solution 1 u simple}) move in the plane $(x,y)$ perpendicularly to each other.

One of two one-line solitons $u^{(1)}$ or $u^{(2)}$ (not both) of superposition (\ref{VN solution 1 u simple}) due to (\ref{VN two-solution 1 varphi1Normals}) and (\ref{Parametrization}) can be \,"stopped", i. e. by special choice of spectral parameter $\lambda_1$ one can take ones of velocities $V_{1}=0$ or $V_{2}=0$ equal to zero. For example one can choose $V_{2}=0$, this achieves by the use of
(\ref{VN two-solution 1 varphi1Normals}) and (\ref{Parametrization}) for $\lambda_1$ satisfying to condition
\begin{equation}\label{SplittingCondition2VN[K,0]}
\kappa\lambda_1^3+{\overline{\kappa}}{\overline{\lambda_1^3}}=0.
\end{equation}

It was shown in the papers~\cite{DubrTopBasTMP1,DubrTopBas_arxiv} that the limiting procedure for calculation of exact solutions $u$ of VN equation with zero asymptotic values at infinity $u|_{|z|^2\rightarrow\infty}=-\epsilon\rightarrow0$ ($\overline\partial$-dressing on zero energy level) can be defined by the following way
\begin{equation}\label{ZeroAsymptotic}
\epsilon\rightarrow0,\quad\mu_n\rightarrow0,\quad \frac{\epsilon}{\mu_n}\rightarrow-\overline{\lambda}_n\neq0,\quad n=1,\ldots,N.
\end{equation}
It is assumed that under procedure (\ref{ZeroAsymptotic}) the relations $\lambda_n=i\tau_n\lambda_1$ from
(\ref{Parametrization}) remain to be valid.

In the limit (\ref{ZeroAsymptotic}) two-line soliton solution
(\ref{VN solution 1 u simple}) converts to linear superposition
\begin{equation}\label{VSchrE0Limit}
u(x,y,t)=u^{(1)}_{\epsilon=0}+u^{(2)}_{\epsilon=0}=
\sum\limits_{n=1}^{2}\frac{|\lambda_n|^2}{2\cosh^2{\frac{\tilde\varphi_{n}(x,y,t) + \phi_{0n}}{2}}}
\end{equation}
of two one-line solitons $u^{(1)}_{\epsilon=0}$ and $u^{(2)}_{\epsilon=0}$
\begin{equation}\label{OneSoliton[k0]E0}
u^{(n)}_{\epsilon=0}(x,y,t)=\frac{|\lambda_{n}|^2}{2\cosh^2{\frac{\tilde\varphi_{n}(x,y,t) + \phi_{0n}}{2}}},\quad n=1,2
\end{equation}
with phases $\tilde\varphi_{n}(x,y,t)$ given due to (\ref{VN two-solution 1 varphi1}) and (\ref{ZeroAsymptotic}) by formulas
\begin{equation}\label{phasesVN[2,0]}
\tilde\varphi_{n}(x,y,t)=2|\lambda_n|(\vec{N}_n\vec{r}-V_n t),\quad \vec{N}_n=\left(\frac{\lambda_{nI}}{|\lambda_{n}|},\frac{\lambda_{nR}}{|\lambda_{n}|}\right),\quad n=1,2.
\end{equation}
Here $\vec{r}=(x,y)$,  $\vec{N}_n$ are unit vectors of normals to lines of constant values of phases $\tilde\varphi_{n}(x,y,t)$;  $V_n$ are corresponding velocities of one-line solitons
\begin{equation}\label{velocityVN[2,0]}
V_1=-\frac{\mbox{Im}(\kappa\lambda_1^3)}{|\lambda_1|},\quad
V_2=-\frac{\mbox{Im}(\kappa\lambda_2^3)}{|\lambda_2|}=\frac{\tau_2^3\mbox{Re}(\kappa\lambda_1^3)}{|\tau_2||\lambda_1|}
\end{equation}
derived by the use of (\ref{VN two-solution 1 varphi1Normals}), (\ref{Parametrization}) and (\ref{ZeroAsymptotic}).
By special choice of spectral parameter $\lambda_1$ one of two one-line solitons $u^{(1)}_{\epsilon=0}$ or $u^{(2)}_{\epsilon=0}$ (not both) in linear superposition (\ref{VSchrE0Limit}) can be \,"stopped".

The solution in the form of another nonlinear superposition of $N\geqslant2$ one-line solitons  (\ref{VN solution 1 u}) is given by (\ref{MultiSolitonGF}) with parameters $a_n$, $(\mu_n,\lambda_n)$ and $s_n$ satisfying to (\ref{[1,2,0]VNSolitons}), (\ref{p_kVNElliptic1}), the conditions (\ref{SplittingCondition1}) or (\ref{Parametrization}) and (\ref{SplittingCondition2}) also must be fulfilled. The  condition (\ref{SplittingCondition2}) due to (\ref{[1,2,0]VNSolitons}) leads to relation
\begin{equation}\label{cond1}
\kappa\lambda_1^3+\overline{\kappa}\overline{\lambda_1^3}=0,
\end{equation}
so the  solution  (\ref{MultiSolitonGF})   takes the following real form
\begin{equation}\label{MultiSoliton[k0]E}
u(x,y,t)=-\epsilon+\frac{|\lambda_{1}-\mu_{1}|^2}{2\cosh^2{\frac{\varphi_{1}(x,y,t) + \phi_{01}}{2}}} +\sum\limits_{n=2}^{N}\frac{|\lambda_{n}-\mu_{n}|^2}{2\cosh^2{\frac{\varphi_{n}(x,y) + \phi_{0n}}{2}}}
\end{equation}
with phases $\varphi_{n}$ (\ref{Phazes1NsolitonGF}), (\ref{PhazesnNsolitonGF}) are given due to
(\ref{[1,2,0]VNSolitons}), (\ref{p_kVNElliptic1}) and (\ref{Parametrization}), (\ref{cond1})  by expressions
\begin{equation}\label{PhasesVN[K,0]}
\varphi_1(x,y,t)= 2|\lambda_1|\left(1+\frac{\epsilon}{|\lambda_1|^2}\right)(\vec{N}_1\vec{r}-V_1t),\quad
\varphi_n(x,y)= 2|\lambda_1|\left(\tau_n+\frac{\epsilon}{\tau_n|\lambda_1|^2}\right)(\vec{N}_2\vec{r}).
\end{equation}
In formulas (\ref{PhasesVN[K,0]}) $n=2,\ldots,N$, $\vec{r}=(x,y)$; the unit vectors of normals $\vec{N}_{1}$, $\vec{N}_{2}$ and velocity $V_1$ are given by following formulas
\begin{equation}\label{normalsVN[K,0]}
\vec{N}_1=\left(\frac{\lambda_{1I}}{|\lambda_{1}|},\frac{\lambda_{1R}}{|\lambda_{1}|}\right),\quad \vec{N}_2=\left(\frac{\lambda_{1R}}{|\lambda_{1}|},-\frac{\lambda_{1I}}{|\lambda_{1}|}\right),\quad
V_{1}=\frac{i\kappa\lambda^3_1}{|\lambda_1|}
\left(1+\frac{\epsilon\left(\epsilon-|\lambda_1|^2\right)}{|\lambda_1|^4}\right).
\end{equation}
One-line soliton $u^{(1)}(x,y,t)=-\epsilon+\tilde{u}^{(1)}(x,y,t)$ of nonlinear superposition (\ref{MultiSoliton[k0]E})  due to (\ref{PhasesVN[K,0]}) and (\ref{normalsVN[K,0]})   moves in the plane $(x,y)$ perpendicularly to others stationary solitons $u^{(n)}(x,y)=-\epsilon+\tilde{u}^{(n)}(x,y)$, $(n=2,\ldots,N)$  of this superposition with parallel lines of constant values of  phases $\varphi_n(x,y)$.
Evidently particular case of (\ref{VN solution 1 u simple}) with $V_2=0$ due to
(\ref{SplittingCondition2VN[K,0]}) and (\ref{cond1}) coincides with  two-line soliton (for $N=2$) nonlinear superposition (\ref{MultiSoliton[k0]E}).

In the limit $\epsilon\rightarrow0$ following to the rules (\ref{ZeroAsymptotic}), under requirement
that relations $\lambda_n=i\tau_n\lambda_1$ from
(\ref{Parametrization}) and (\ref{cond1}) remain to be valid, we obtain from (\ref{MultiSoliton[k0]E}) linear superposition of $N$ one-line solitons
\begin{equation}\label{MultiSoliton[k0]E0}
 u=u_{\epsilon=0}^{(1)}(x,y,t)+\sum\limits_{n=2}^{N}u_{\epsilon=0}^{(n)}(x,y)
=\frac{|\lambda_{1}|^2}
{2\cosh^2{\frac{\tilde\varphi_{1}(x,y,t) + \phi_{01}}{2}}} +\sum\limits_{n=2}^{N}\frac{|\lambda_{n}|^2}{2\cosh^2{\frac{\tilde\varphi_{n}(x,y) + \phi_{0n}}{2}}},
\end{equation}
here the phases $\tilde\varphi_{n}$ given by expressions
\begin{equation}\label{phasesVN[K,0]E0}
\tilde\varphi_{1}(x,y,t)=2|\lambda_1|\left(\vec{N}_1\vec{r}-\frac{i\kappa\lambda^3_1}{|\lambda_1|} t\right),\quad \tilde\varphi_{n}(x,y)=2\tau_n|\lambda_1|(\vec{N}_2\vec{r}),\quad n=2,\ldots,N
\end{equation}
are obtained from phases   $\varphi_{n}$ (\ref{PhasesVN[K,0]}) in
limit $\epsilon\rightarrow0$ (\ref{ZeroAsymptotic}).
The first one-line soliton $u^{(1)}_{\epsilon=0}(x,y,t)$ from linear superposition (\ref{MultiSoliton[k0]E0}) moves with velocity $V_1=\frac{i\kappa\lambda^3_1}{|\lambda_1|}$ in the plane $(x,y)$  perpendicularly to other $(N-1)$  stationary one-line solitons
 $u^{(n)}_{\epsilon=0}(x,y)$.

It is easy to show  that the subsums of arbitrary numbers of solitons $u^{(n)}_{\epsilon=0}$, $(n=1,\ldots,N)$ from (\ref{MultiSoliton[k0]E0}) are also solutions of VN equation. So the set of such solutions can be divided in two subsets: subset of nonstationary linear superpositions (with the first moving line soliton  $u^{(1)}_{\epsilon=0}(x,y,t)$ in the sum) of line solitons and subset of stationary linear superpositions (without moving line soliton $u^{(1)}_{\epsilon=0}(x,y,t)$ in the sum) of stationary line solitons.

\section{Linear superpositions of plane wave type periodic solutions for Veselov-Novikov equation}

Multi-line soliton solutions are calculated in preceding section by imposing reality condition $u=\bar{u}$ on complex solutions (\ref{uN=1Gen}), (\ref{uTwoSolGen}), (\ref{MultiSolitonGF}) and (\ref{SubSystSuperp}) with additional assumption of real phases $\Delta F(\mu_n,\lambda_n)= \overline{\Delta F(\mu_n,\lambda_n)}$   (\ref{s_k,Delta F}). In contrast the application of reality condition $u=\bar{u}$ with assumption of pure imaginary phases $\Delta F(\mu_n,\lambda_n)=-\overline{\Delta F(\mu_n,\lambda_n)}$  (\ref{s_k,Delta F})  leads to plane wave type periodic solutions and their's superpositions.

Plane wave type solutions can be obtained by this way for example by the following choice of parameters $a_n,(\mu_n,\lambda_n)$ and $s_n$ in
(\ref{sum delta_kernel_satisfied_potent}) -- (\ref{SubSystSuperp})~\cite{DubrTopBasTMP1,DubrTopBas_arxiv}:
\begin{equation}\label{[1,2,0]VNSolitons ps}
 \mu_n = \frac{\epsilon}{\overline{\lambda}_n}, \quad
a_n=\left|\frac{\lambda_n-\mu_n}{\lambda_n+\mu_n}\right|e^{i\arg{a_n}}, \quad
s_n=- i e^{i\arg{a_n}}\sign\left({\frac{\lambda_n-\mu_n}{\lambda_n+\mu_n}}\right),\quad n=1,\ldots,N.
\end{equation}

Simple plane wave type periodic solution corresponding to one arbitrary pair of spectral variables $(\mu_n,\lambda_n)$, $(n=1,\ldots, N)$, due to (\ref{uN=1Gen}) and (\ref{[1,2,0]VNSolitons ps})  takes the form
\begin{equation}\label{VN solution 1 u ps}
 u^{(n)}(x,y,t) = -\epsilon +\tilde u^{(n)}(x,y,t)= -\epsilon -
  \frac{|\lambda_n - \mu_n|^2}{2\cos^2\left(\frac{\varphi_n(x,y,t) + \arg{a_n}}{2}\mp\frac{\pi}{4}\right)},
\end{equation}
where sign "$-$"\, correspond to case of $|\lambda_n|>|\mu_n|$ and sign "$+$"\, -- case of $|\lambda_n|<|\mu_n|$.
The real phases $\varphi_n(x,y,t):=-i\Delta F(\mu_n,\lambda_n)$ in (\ref{VN solution 1 u ps}) due to (\ref{s_k,Delta F}) and (\ref{[1,2,0]VNSolitons ps}) are given by expressions
\begin{equation}\label{VN two-solution 1 varphi1 ps}
\varphi_n(x,y,t)= 2|\lambda_n|\left(\frac{\epsilon}{|\lambda_n|^2}-1\right)(\vec{N}_n\vec{r}-V_n t),\quad n=1,\ldots,N.
\end{equation}
In (\ref{VN solution 1 u ps}), (\ref{VN two-solution 1 varphi1 ps}) $\vec{r}=(x,y)$;  $\vec{N}_n$ are unit vectors of normals to lines of constant values of phases $\varphi_{n}(x,y,t)$ and velocities $V_n$  of periodic solutions are given by expressions:
\begin{equation}\label{VN two-solution 1 varphi1Normals ps}
\vec{N}_n=\left(\frac{\lambda_{nR}}{|\lambda_{n}|},-\frac{\lambda_{nI}}{|\lambda_{n}|}\right),\quad
V_{n}=-\frac{1}{|\lambda_n|}
\left(1+\frac{\epsilon\left(\epsilon+|\lambda_n|^2\right)}{|\lambda_n|^4}\right)\mathrm{Re}(\kappa\lambda^3_n),\quad
 n=1,\ldots,N.
\end{equation}

Using general formulas (\ref{uTwoSolGen}) and (\ref{MultiSolitonGF}) we also construct nonlinear superpositions of simple wave type periodic solutions of the type
(\ref{VN solution 1 u ps}). The conditions (\ref{SplittingCondition1}) for discrete spectral parameters $(\mu_n,\lambda_n)$, $(n>1)$  in nonlinear superpositions (\ref{uTwoSolGen}) and (\ref{MultiSolitonGF})
due to (\ref{[1,2,0]VNSolitons ps}) in considered
case lead to following parametrization of $(\mu_n,\lambda_n)$
\begin{equation}\label{Parametrization ps}
\lambda_n=i\tau_{n}\lambda_1, \quad \mu_n=i\tau_{n}^{-1}\mu_1,\quad n=2,\ldots,N
\end{equation}
with arbitrary real constants $\tau_n$.
Nonlinear superposition (\ref{uTwoSolGen}) of two simple plane wave type periodic solutions of the type (\ref{VN solution 1 u ps}) due to (\ref{[1,2,0]VNSolitons ps}) and (\ref{Parametrization ps}) takes the form
\begin{equation}\label{VN solution 1 u simple ps}
 u(x,y,t)= -\epsilon+\sum\limits_{n=1}^{2}\tilde u^{(n)}(x,y,t)=-\epsilon-\sum\limits_{n=1}^{2}
\frac{|\lambda_n - \mu_n|^2}{2\cos^2\left({\frac{\varphi_{n}(x,y,t) + \arg{a_n}}{2}}\mp\frac{\pi}{4}\right)},
\end{equation}
where sign "$-$"\, correspond to case of $|\lambda_n|>|\mu_n|$ and sign "$+$"\, -- case of $|\lambda_n|<|\mu_n|$.
The phases $\varphi_{n}(x,y,t)$ in solution (\ref{VN solution 1 u simple ps}) are given by
(\ref{VN two-solution 1 varphi1 ps}) with parametrization (\ref{Parametrization ps}).
Due to expressions for vectors of normals (\ref{VN two-solution 1 varphi1Normals ps}) and to parametrization (\ref{Parametrization ps}) it is evident that lines of constant values of phases $\varphi_{n}(x,y,t), (n=1,2)$ for solutions $u^{(1)}=-\epsilon+\tilde u^{(1)}$ and $u^{(2)}=-\epsilon+\tilde u^{(2)}$ of superposition (\ref{VN solution 1 u simple ps}) move perpendicularly to each other.

One of two simple plane wave type periodic solutions $u^{(1)}$ or $u^{(2)}$ (not both) of superposition (\ref{VN solution 1 u simple ps}) due to (\ref{VN two-solution 1 varphi1Normals ps}) and (\ref{Parametrization ps}) can be made stationary, i. e. by special choice of spectral parameter $\lambda_1$ one can take ones of velocities $V_{1}=0$ or $V_{2}=0$ equal to zero.
For example one can choose $V_2=0$, this achieves due to (\ref{VN two-solution 1 varphi1Normals ps}) and (\ref{Parametrization ps}) for $\lambda_1$ satisfying to condition
\begin{equation}\label{SplittingCondition2VN[K,0] ps}
\kappa\lambda_1^3-{\overline{\kappa}}{\overline{\lambda_1}^3}=0.
\end{equation}

It was shown in the papers~\cite{DubrTopBasTMP1,DubrTopBas_arxiv} that
the limiting procedure of calculation of exact solutions $u$ of VN
equation with zero  values of parameter $\epsilon=0$
($\overline\partial$-dressing on zero energy level)
can be defined by the following way
\begin{equation}\label{ZeroAsymptotic ps}
\epsilon\rightarrow0,\quad\mu_n\rightarrow0,\quad \frac{\epsilon}{\mu_n}\rightarrow\overline{\lambda}_n\neq0,\quad n=1,\ldots,N.
\end{equation}
Such procedure is applicable also in considered case of plane wave type solutions and their's superpositions.
It is assumed that under procedure (\ref{ZeroAsymptotic ps}) the relations $\lambda_n=i\tau_n\lambda_1$ from
(\ref{Parametrization ps}) remain to be valid.

In the limit (\ref{ZeroAsymptotic ps}) nonlinear superposition (\ref{VN solution 1 u simple ps}) of two plane wave type periodic solutions
 converts to linear superposition
\begin{equation}\label{VSchrE0Limit ps}
u(x,y,t)=u^{(1)}_{\epsilon=0}+u^{(2)}_{\epsilon=0}=
-\sum\limits_{n=1}^{2}\frac{|\lambda_n|^2}{2\cos^2\left({\frac{\tilde\varphi_{n}(x,y,t) + \arg{a_n}}{2}}-\frac{\pi}{4}\right)}
\end{equation}
of two periodic solitons $u^{(1)}_{\epsilon=0}$ and $u^{(2)}_{\epsilon=0}$
\begin{equation}\label{OneSoliton[k0]E0 ps}
u^{(n)}_{\epsilon=0}(x,y,t)=-\frac{|\lambda_{n}|^2}{2\cos^2\left({\frac{\tilde\varphi_{n}(x,y,t) + \arg{a_n}}{2}}-\frac{\pi}{4}\right)},\quad n=1,2,
\end{equation}
 the phases $\tilde\varphi_{n}(x,y,t)$ in (\ref{VSchrE0Limit ps}), (\ref{OneSoliton[k0]E0 ps})   are given due to
(\ref{VN two-solution 1 varphi1 ps}) and (\ref{ZeroAsymptotic ps}) by formulas
\begin{equation}\label{phasesVN[2,0] ps}
\tilde\varphi_{n}(x,y,t)=-2|\lambda_n|(\vec{N}_n\vec{r}-V_n t),\quad \vec{N}_n=\left(\frac{\lambda_{nR}}{|\lambda_{n}|},-\frac{\lambda_{nI}}{|\lambda_{n}|}\right),\quad n=1,2,
\end{equation}
here $\vec{r}=(x,y)$; $\vec{N}_n$ are unit vectors of normals to lines of constant values of phases $\tilde\varphi_{n}(x,y,t)$.
The corresponding velocities $V_n$  of simple plane wave type periodic solutions (\ref{OneSoliton[k0]E0 ps})
\begin{equation}\label{velocityVN[2,0] ps}
V_1=-\frac{\mbox{Re}(\kappa\lambda_1^3)}{|\lambda_1|},\quad
V_2=-\frac{\mbox{Re}(\kappa\lambda_2^3)}{|\lambda_2|}=\frac{\tau_2^3\mbox{Im}(\kappa\lambda_1^3)}{|\tau_2||\lambda_1|}
\end{equation}
are derived by the use of
(\ref{VN two-solution 1 varphi1Normals ps}),
(\ref{Parametrization ps}) and (\ref{ZeroAsymptotic ps}).
By special choice of spectral parameter $\lambda_1$ one of two of these
periodic solutions,  $u^{(1)}_{\epsilon=0}$ or $u^{(2)}_{\epsilon=0}$
(not both), in linear superposition (\ref{VSchrE0Limit ps}) can be made stationary.

The solution in the form of another nonlinear superposition of $N\geqslant2$ simple plane wave type periodic solutions (\ref{VN solution 1 u ps}) is given by (\ref{MultiSolitonGF}) with parameters $a_n$, $(\mu_n,\lambda_n)$ and $s_n$ satisfying to (\ref{[1,2,0]VNSolitons ps}), the conditions (\ref{SplittingCondition1}) or (\ref{Parametrization ps}) and (\ref{SplittingCondition2}) also must be fulfilled. The condition (\ref{SplittingCondition2}) due to (\ref{[1,2,0]VNSolitons ps}) transforms into following
\begin{equation}\label{cond1 ps}
\kappa\lambda_1^3-\overline{\kappa}\overline{\lambda_1^3}=0,
\end{equation}
so the  solution (\ref{MultiSolitonGF}) takes the form
\begin{equation}\label{MultiSoliton[k0]E ps}
 u(x,y,t)=-\epsilon-\frac{|\lambda_{1}-\mu_{1}|^2}{2\cos^2\left({\frac{\varphi_{1}(x,y,t) + \arg{a_1}}{2}}\mp\frac{\pi}{4}\right)} -\sum\limits_{n=2}^{N}\frac{|\lambda_{n}-\mu_{n}|^2}{2\cos^2\left({\frac{\varphi_{n}(x,y) + \arg{a_n}}{2}}\mp\frac{\pi}{4}\right)},
\end{equation}
where sign "$-$"\, correspond to case of $|\lambda_n|>|\mu_n|$ and sign "$+$"\, -- case of $|\lambda_n|<|\mu_n|$, $n=1,\ldots,N$.
The phases $\varphi_{n}$ (\ref{Phazes1NsolitonGF}), (\ref{PhazesnNsolitonGF}) in superposition (\ref{MultiSoliton[k0]E ps}) due to (\ref{[1,2,0]VNSolitons ps}) and (\ref{Parametrization ps}),
(\ref{cond1 ps}) are given by expressions
\begin{equation}\label{PhasesVN[K,0] ps}
\varphi_1(x,y,t)= 2|\lambda_1|\left(\frac{\epsilon}{|\lambda_1|^2}-1\right)(\vec{N}_1\vec{r}-V_1t),\quad
\varphi_n(x,y)= 2|\lambda_1|\left(\frac{\epsilon}{\tau_n|\lambda_1|^2}-\tau_n\right)(\vec{N}_2\vec{r}),
\end{equation}
where $n=2,\ldots,N$, $\vec{r}=(x,y)$;  unit vectors $\vec{N}_n$ and velocity $V_1$ are given by following formulas
\begin{equation}\label{normalsVN[K,0] ps}
\vec{N}_1=\left(\frac{\lambda_{1R}}{|\lambda_{1}|},-\frac{\lambda_{1I}}{|\lambda_{1}|}\right), \quad \vec{N}_2=\left(-\frac{\lambda_{1I}}{|\lambda_{1}|},-\frac{\lambda_{1R}}{|\lambda_{1}|}\right),\quad
V_{1}=\frac{\kappa\lambda^3_1}{|\lambda_1|}
\left(1+\frac{\epsilon\left(\epsilon+|\lambda_1|^2\right)}{|\lambda_1|^4}\right).
\end{equation}
The lines of constant values of phase $\varphi_1(x,y,t)$ of simple
periodic solution $u^{(1)}(x,y,t)=-\epsilon+\tilde u^{(1)}(x,y,t)$ of
superposition (\ref{MultiSoliton[k0]E ps}) move in plane $(x,y)$
perpendicularly to parallel lines of constant phases $\varphi_n(x,y)$ of
others stationary periodic solutions
$u^{(n)}(x,y,t)=-\epsilon+\tilde u^{(n)}(x,y,t)$, $(n=2,\ldots,N)$ of this superposition.
Evidently particular case of (\ref{VN solution 1 u simple ps}) with $V_2=0$ coincides due to
(\ref{SplittingCondition2VN[K,0] ps}) and (\ref{cond1 ps}) with the case $N=2$ of linear superposition (\ref{MultiSoliton[k0]E ps}).

In the limit $\epsilon\rightarrow0$ ($\overline\partial$-dressing on zero energy level) following to the rules (\ref{ZeroAsymptotic ps}), with assumption that relations $\lambda_n=i\tau_n\lambda_1$ from
(\ref{Parametrization ps}) and (\ref{cond1 ps}) remain to be valid, we obtain from (\ref{MultiSoliton[k0]E ps}) linear superposition of $N$ simple plane wave type periodic solutions
\begin{equation}\label{MultiSoliton[k0]E0 ps}
u=u_{\epsilon=0}^{(1)}(x,y,t)+\sum\limits_{n=2}^{N}u_{\epsilon=0}^{(n)}(x,y)=
\frac{|\lambda_{1}|^2}{2\cos^2\left({\frac{\tilde\varphi_{1}(x,y,t) + \arg{a_1}}{2}}-\frac{\pi}{4}\right)} +\sum\limits_{n=2}^{N}\frac{|\lambda_{n}|^2}{2\cos^2\left({\frac{\tilde\varphi_{n}(x,y) + \arg{a_n}}{2}}-\frac{\pi}{4}\right)},
\end{equation}
here phases $\tilde\varphi_{n}$ are obtained from phases $\varphi_{n}$   (\ref{PhasesVN[K,0] ps}) by setting $\epsilon=0$ in accordance with (\ref{ZeroAsymptotic ps}), these phases have the forms:
\begin{equation}\label{phasesVN[K,0]E0 ps}
\tilde\varphi_{1}(x,y,t)=-2|\lambda_1|
\left(\vec{N}_1\vec{r}-\frac{\kappa\lambda^3_1}{|\lambda_1|}t\right),\quad \tilde\varphi_{n}(x,y)=-2\tau_n|\lambda_1|\left(\vec{N}_2\vec{r}\right),\quad n=2,\ldots,N.
\end{equation}

The lines of constant values of phase $\varphi_{1}(x,y,t)$ of the first periodic solution $u^{(1)}_{\epsilon=0}$ from linear superposition (\ref{MultiSoliton[k0]E0 ps}) move with velocity
$V_1=\frac{\kappa\lambda^3_1}{|\lambda_1|}$  perpendicularly
to lines of constant phases $\varphi_{n}(x,y)$ of $(N-1)$
other stationary periodic solutions $u^{(n)}_{\epsilon=0},
(n=2,\ldots,N)$ of this superposition.

It is easy to show that the subsums of arbitrary numbers
of solutions $u^{(n)}_{\epsilon=0}$, $(n=1,\ldots,N)$
from (\ref{MultiSoliton[k0]E0 ps}) are also solutions of VN equation.
So the set of constructed in present section  solutions can be divided in two subsets: subset of nonstationary linear superpositions (with the first moving line soliton  $u^{(1)}_{\epsilon=0}(x,y,t)$ in the sum) and subset of stationary linear superpositions (without moving line soliton $u^{(1)}_{\epsilon=0}(x,y,t)$ in the sum).

\section{Appendix}

In appendix some statements of the paper are proved.   
The VN equation (\ref{VN}) can be represented in the form:
\begin{equation}\label{Appendix1}
\mathrm{N}(u)=\mathrm{L}(u)+\mathrm{Bil}(u,u),
\end{equation}
with linear part
\begin{equation}\label{Appendix11}
\mathrm{L}(u)=u_{t} + \kappa u_{zzz} + \overline \kappa u_{\bar z\bar z\bar z};
\end{equation}
and nonlinear part
\begin{equation}\label{Appendix2}
\mathrm{Bil}(u,u)=3\kappa(u\partial^{-1}_{\bar z}u_{z})_{z}+3\bar\kappa(u\partial^{-1}_{z}u_{\bar z})_{\bar z}.
\end{equation}
in terms of bilinear form
\begin{equation}\label{Appendix21}
\mathrm{Bil}(u,v)=3\kappa(u\partial^{-1}_{\bar z}v_{z})_{z}+3\bar\kappa(u\partial^{-1}_{z}v_{\bar z})_{\bar z}.
\end{equation}
The proof of the fact that (\ref{MultiSolitonGF}) is solution of VN equation can be done by direct substitution
(\ref{MultiSolitonGF}) into (\ref{Appendix1}) and regrouping arising terms:
\begin{equation}\label{Appendix3}
\mathrm{N}\left(-\epsilon+\sum\limits_{k=1}^{N}\tilde u^{(k)}\right)=\sum\limits_{k=1}^{N}\mathrm{N}(u^{(k)})+
\sum\limits_{k\neq j}\mathrm{Bil}\left(\tilde u^{(k)},\tilde u^{(j)}\right).
\end{equation}
Here due to (\ref{MultiSolitonGF}) $u^{(1)}(x,y,t)=\tilde u^{(1)}(x,y,t)-\epsilon$ are nonstationary and $u^{(k)}(x,y)=\tilde u^{(k)}(x,y)-\epsilon$, $k=2,\ldots,N$ stationary solutions of VN equation with corresponding phases (\ref{Phazes1NsolitonGF}) and
(\ref{PhazesnNsolitonGF}).

Using formula (\ref{uN=1Gen}), taking into account (\ref{Phazes1NsolitonGF}) and
(\ref{PhazesnNsolitonGF}) and the relation between derivatives of phases
\begin{equation}\label{Appendix4}
\frac{\partial\varphi_k}{\partial \bar z}=\frac{\epsilon}{\lambda_k \mu_k}\frac{\partial\varphi_k}{\partial  z},\quad k=1,\ldots,N
\end{equation}
one obtains due to (\ref{SplittingCondition1})
\begin{equation}\label{Appendix5}
\mathrm{Bil}(\tilde u^{(1)},\tilde u^{(k)})=\frac{3\kappa}{\epsilon}(\lambda_k\mu_k+\lambda_1\mu_1)(\tilde u^{(1)}\tilde u^{(k)})_z+3\bar\kappa\epsilon\left(\frac{1}{\lambda_k\mu_k}+\frac{1}{\lambda_1\mu_1}\right)(\tilde u^{(1)}\tilde u^{(k)})_{\bar z}=0,\quad k=2,\ldots,N.
\end{equation}
Quite analogously one calculates for $k\neq m$ ($k,m=2,\ldots,N$)
\begin{eqnarray}\label{Appendix6}
\mathrm{Bil}(\tilde u^{(k)},\tilde u^{(m)})=3\Bigg(\frac{\bar{\kappa} \epsilon^2}{\lambda^2_k\mu^2_k}+
\frac{\bar{\kappa} \epsilon^2}{\lambda_k\mu_k \lambda_m\mu_m}+\frac{\kappa\lambda_k\mu_k}{\epsilon}+\frac{\kappa\lambda_m\mu_m}{\epsilon}\Bigg)(\tilde u^{(m)}\tilde u_{z}^{(k)})+\nonumber\\
3\Bigg(\frac{\bar{\kappa} \epsilon^2}{\lambda^2_m\mu^2_m}+
\frac{\bar{\kappa} \epsilon^2}{\lambda_k\mu_k \lambda_m\mu_m}+\frac{\kappa\lambda_k\mu_k}{\epsilon}+\frac{\kappa\lambda_m\mu_m}{\epsilon}\Bigg)(\tilde u_{z}^{(m)}\tilde u^{(k)})=
6\Bigg(\frac{\bar{\kappa}\epsilon^2}{\lambda^2_{1}\mu^2_1}-\frac{\kappa\lambda_1\mu_1}{\epsilon}\Bigg)(\tilde u^{(k)}\tilde u^{(m)})_{z}=0
\end{eqnarray}
here in derivation of (\ref{Appendix6}) the role of constraints (\ref{SplittingCondition1}) and (\ref{SplittingCondition2}) on parameters $(\lambda_1,\mu_1)$ and $(\lambda_k,\mu_k)$, $k=2,\ldots,N$ is clear.

Due to (\ref{Appendix3}), (\ref{Appendix5}) and (\ref{Appendix6}) one derives the required  result for nonlinear superposition (\ref{MultiSolitonGF})
\begin{equation}\label{Appendix7}
u(x,y,t)=-\epsilon+\tilde u^{(1)}(x,y,t)+\sum\limits_{k=2}^{N}\tilde u^{(k)}(x,y)=u^{(1)}(x,y,t)+\sum\limits_{k=2}^{N}u^{(k)}(x,y)+(N-1)\epsilon.
\end{equation}
This superposition due to (\ref{Appendix3}), (\ref{Appendix5}) and (\ref{Appendix6}) is the solution of VN equation
\begin{equation}\label{Appendix8}
\mathrm{N}(u)=\mathrm{N}(-\epsilon+\sum\limits_{k=1}^{N}\tilde u^{(k)})=\sum\limits_{k=1}^{N}\mathrm{N}(u^{(k)})=0.
\end{equation}
Nonlinear superposition (\ref{Appendix7}) modulo $(N-1)\epsilon$ is almost linear superposition of solutions $u_{(k)}$ -- the sum of $u_{(k)}$.

The statement expressed by formulas (\ref{SubSystSuperp}), (\ref{MultiSoliton[k0]E}) and (\ref{MultiSoliton[k0]E0}) are follow from (\ref{Appendix8}). Solutions (\ref{SubSystSuperp}) is particular case of (\ref{MultiSolitonGF}); solution (\ref{MultiSoliton[k0]E}) is specialization of (\ref{MultiSolitonGF}) to real solution $u$; solution (\ref{MultiSoliton[k0]E0}) is zero limit $\epsilon\rightarrow0$ of  (\ref{Appendix7}), (\ref{Appendix8}).

Quite analogously to solitonic case corresponding proof for exact periodic solutions (\ref{MultiSoliton[k0]E ps}) and (\ref{MultiSoliton[k0]E0 ps}) can be done.

\section{Acknowledgements}
This research work is supported: 1. by scientific Grant of fundamental researches of Novosibirsk State Technical University  (2011); 2. by the Grant (registration number 2.1.1/1958) of Ministry of Science and Education of Russia Federation via analytical departmental special programm "Development of potential of High School" (2011).

\end{document}